%% file: paper.tex
\documentstyle[twoside,fleqn,espcrc2,epsf]{article}
\RequirePackage[hyperindex]{hyperref}
\let\figcaption\caption

\def\MSbar{{\ifmmode\let\finish=\relax\else$\let\finish=$\fi
            \mathpalette{\hbox\bgroup$}{\overline{MS}\egroup$}%
            \finish}}
 
\newcommand\etal{{\it et al.}}
\newcommand\bicg{\hbox{$ BiCG\gamma_5 $}}
\newcommand\bicgstab{\hbox{$ BiCGstab $}}
\newcommand\forcefill{\aftergroup\hfill}

\begin{document}
 
\title{Comparison of Inversion Algorithms for Wilson Fermions on the CM5\thanks{%
Talk presented by Rajan Gupta at the workshop ``Accelerating Fermion
Algorithms'', J\"ulich, Feb 1996.  The calculations reported here have
been done on the CM5 at LANL as part of the DOE HPCC Grand Challenge
program, and at NCSA under a Metacenter allocation.}}
 
\author{Rajan Gupta,%
        \address{T-8 Group, MS B285, Los Alamos National
        Laboratory, Los Alamos, New Mexico 87545 U.~S.~A.~}
        \addtocounter{address}{-1}%
        Tanmoy Bhattacharya,\addressmark\ %
        and Gregory Kilcup%
        \address{Physics Department, The Ohio State University, Columbus, 
        OH 43210 U.~S.~A.~}
}
 
\begin{abstract}
 
This talk presents results of a comparitive study of iterative
algorithms like minimal residue ($MR$) and conjugate gradient ($CG$,
$BiCG\gamma_5$, and \bicgstab) used for inverting the Dirac matrix
$M$.  The tests were done on the Connection Machine CM-5 using $32^3
\times 64$ lattices. The fermion action used is of Wilson type, both
with and without the clover term.  The overall conclusion is that
preconditioned over-relaxed $MR$ is the simplest, uses the least
memory, and is comparable in performance to \bicgstab.  We find these
two algorithms to be equally robust, $i.e.$ insensitive to the starting
solution and to round-off errors.

\end{abstract}
 
\maketitle
 
%
%
%

\section{INTRODUCTION}
\label{s_intro}

The solution of the Dirac equation for an arbitrary source constitutes
the dominant fraction of computer time used in the numerical
simulation of both quenched and unquenched QCD.  In quenched
simulations, the Dirac propagator is used in the construction of
correlation functions of observables involving quark fields. Depending
on how many observables are measured using a given set of quark
propagators, the matrix inversion is $40-70\%$ of the total CPU time.
In full QCD simulations this number jumps to $> 90\%$ as matrix
inversion is also needed in the update.  Large scale simulations
(quenched and unquenched) currently consume the equivalent of many
giga-flop years on a variety of supercomputers. Thus, fast solvers are
the key to progress, and we consider algorithm improvements by $ >
20\%$ significant.  In this talk we catalogue and compare the
efficiency of commonly used algorithms (Minimal Residue($MR$),
conjugate gradient ($CG$, $BiCG\gamma_5$, and \bicgstab)) for Wilson
and Clover class of actions.  Further details of the lattice
generation, the calculation of quark propagators, and the analysis of 
the meson and baryon spectrum are given in Ref.~\cite{rLANLhm95}.

\section{TECHNICAL DETAILS}
\label{s_tech}

The Dirac matrix discretised $a\ la$ Wilson on a 4-dimensional hypercubic 
grid can be written as 
\begin{equation}
M  =  1 + \kappa D(r)
\end{equation}
where the matrix $D(r)$ connects a given site to its 8 nearest
neighbours, $r$ is the Wilson parameter introduced to remove the doubling 
problem, and $\kappa$ is a parameter in terms of which the
quark mass is defined as
\begin{equation}
m_q  = {1\over 2a} \bigg( {1 \over \kappa} -  {1 \over \kappa_c} \bigg)
\end{equation}
where $a$ is the lattice spacing and $\kappa_c$ denotes the chiral limit 
at which the matrix becomes singular. The quark propagator $\chi$ is 
the solution to the set of linear equations 
\begin{equation}
(1 + \kappa D(r)) \chi  =  \phi
\end{equation}
where $\phi$ is an arbitrary source vector. For a delta function
source it represents the propagation of a quark from that point to all
other points on the lattice.

\subsection{\bf Properties of the Dirac matrix} 
\begin{figure}[t]
{\epsfxsize=0.9\hsize\epsfbox{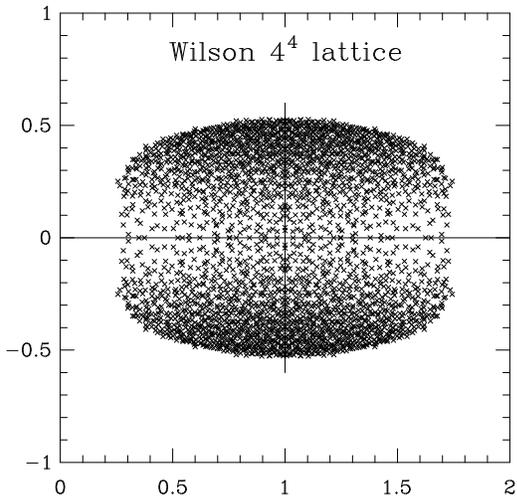}}
\figcaption{Distribution of eigenvalues for a $4^4$ lattice with Wilson 
action in the confined phase at $\beta=5.0$ and $\kappa=0.14$. Data 
provided by A. Borici and P. deForcrand \protect\cite{rBorici}.}
\label{f_borici}
\vskip -0.2 truein
\end{figure}

The Dirac matrix satisfies 
the identity $\gamma_5M\gamma_5=M^\dagger$ which is retained by both
the Wilson and Staggered discretized versions. As a result the
eigenvalues of $M$ come in complex-conjugate pairs.  For the staggered
theory, they lie along a line in the imaginary direction at $m_q$. For
the Wilson theory they are distributed as shown in
Fig.~\ref{f_borici}.  The eigenvalues lying along the real axis in the
confined phase acquire an imaginary part at the chiral/deconfinement
transition. I believe that these real eigenvalues play a central role
in the rate of convergence, and a study of their distribution
and change with preconditioning may provide clues to developing better
algorithms.

\subsection{\bf Convergence Criteria}
We define the remainder at the $n^{th}$ iteration to be $r_n = M\chi_n
- \phi$ and in the figures plot the convergence condition ${\cal C}
\equiv |r|^2 / |\chi|^2 $. On $32^3 \times 64$ lattices, 32 bit IEEE
precision restricts the precision to which ${\cal C}$ can be measured
to $\approx 10^{-14}$ due to round-off errors. In the figures we plot
the iterated residue, so we compare algorithms using the
same preconditioned matrix.  Also, since our final stopping criteria
is with respect to the true residue, a comparison of the total number
of multiplies by the original Dirac matrix $M$ is also meaningful,
irrespective of the preconditioning.

\begin{table}
\caption{Operations per iteration in the various algorithms 
and number of stored vectors.}
\input {t_algops}

\label{t_algops}
\vskip -0.2 truein
\end{table}

\subsection{\bf Polynomial Preconditioning}
The following pre-conditioned systems are used to solve for 
$\chi$~\cite{rALGlanl}
\begin{eqnarray}
(1 - \kappa^2 D^2 ) \chi &{}={}& M'M \chi = (1 - \kappa D) \phi \nonumber \\
(1 - \kappa^4 D^4 ) \chi &{}={}&  (1 + \kappa^2 D^2) (1 - \kappa D)\phi \ ,
\label{eq:MRpre}
\end{eqnarray}
where we define $M' = 1 - \kappa D(r)$.  We label these second and
fourth order pre-conditioned systems by adding subscripts to the
algorithm name.  The second-order form is equivalent to red/black
pre-conditioning and one can halve the number of operations by solving
for only the red (or black) sites and reconstructing the other from
these. On the CM5 the necessary reorganization of data arrays to make
use of this simplification has not been done. The tests have been
performed solving for the full matrix as they do not affect the number
of matrix multiplies needed (the red and black systems are solved
independently and simultaneously).

\subsection{\bf Staggered Fermions} The staggered version of 
the Dirac matrix has the form $m_q + iA$, where $A$ is hermitian. Thus, 
the hermitian product $M^\dagger M = (m_q^2 + A^2)$ is also second
order preconditioned. For such systems $CG$ has proven to be the
optimal algorithm \cite{rbicg5}.\looseness-1

\section{ALGORITHMS}
\label{s_algs}
 
In this section we give the precise implementation of the various
algorithms since the speed of convergence can depend on the details.
We also give the performance characteristics of these algorithms, 
saving a comparitive study for the next section.

\subsection{\bf Minimal Residue}

The minimal residue algorithm consists of the following steps:
\begin{eqnarray}
do\ i=1,&\omit\span\ until\_converged \nonumber \\
    \alpha  &{}={}& {\langle M r_{i-1}, r_{i-1} \rangle / 
                \langle M r_{i-1}, M r_{i-1} \rangle}  \nonumber \\
    \chi_i  &{}={}&  \chi_{i-1} - \omega \ \alpha \ r_{i-1} \nonumber \\
\noalign{\penalty-9000}
     r_i    &{}={}&  r_{i-1}    - \omega \ \alpha \ M \ r_{i-1} \nonumber \\
	    &\omit\span test\ convergence \forcefill\nonumber \\
enddo\forcefill&&
\label{eq:MR}
\end{eqnarray}
where $\alpha$ is a complex number of $O(1)$, and the order of the
hermitian dot product is important. (The algorithm fails to 
converge if we use $\alpha^*$ instead of $\alpha$, and is about $50\%$ 
slower if we use real$(\alpha)$). For initial guess one can
use $\chi_0=0$ or any other trial vector. The only restriction seems to
be that $\langle M r_{i-1}, r_{i-1} \rangle $ should not be small.
This situation arises for $\chi_n$ produced by the $CG$ class of
algorithms.  Thus, $MR$ should {\bf NOT} be used following even a few
$CG$ steps.

\begin{figure}[t]
{\epsfxsize=0.9\hsize\epsfbox{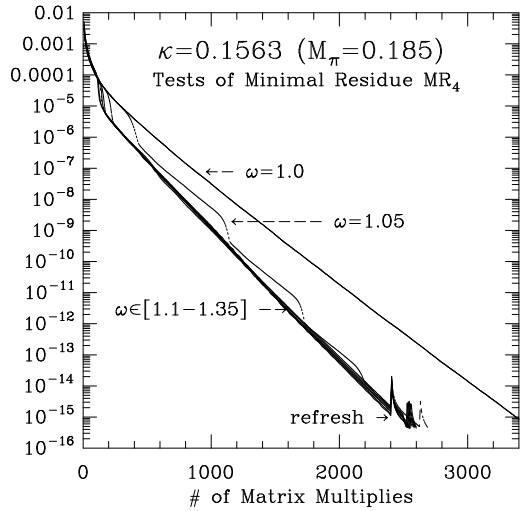}}
\figcaption{Behavior of $MR_4$ versus the over-relaxation parameter 
$\omega$. The band for $1.1 \le \omega \le 1.35$ shows the insensitivity 
of the algorithm to the precise value of $\omega$ in this range. The case
$\omega=1.05$ shows a step like behavior bounded by the ``best'' and
normal ($\omega=1.0$) case. The spikes at the end are due to
refreshing the remainder.} 
\label{f1_MRk3}
\end{figure}

The parameter $\omega$ in Eq.~\ref{eq:MR} is an over-relaxation
parameter.  We find that, to within few percent, the convergence rate
is insensitive to $\omega$ in the range $1.1 - 1.35$, and gives the
envelope of best convergence.  For $1.0 \le \omega \le 1.08$ the
convergence fluctuates between the $\omega=1$ (poorest) and
$\omega=1.1$ (best) cases as\vadjust{\penalty10000} shown in
Fig.~\ref{f1_MRk3}.

We find that higher orders of polynomial pre-conditioning extend the
range of applicability (towards smaller quark masses) of the $MR$
algorithm. There is a $\sim 10\%$ degradation in CPU time with each order, 
$MR_4$ takes roughly $10\%$ more computer time but can be used for 
lighter quark masses \cite{rALGlanl}. 

\subsection{\bf Conjugate Gradient}

\begin{figure}[t]
{\epsfxsize=0.9\hsize\epsfbox{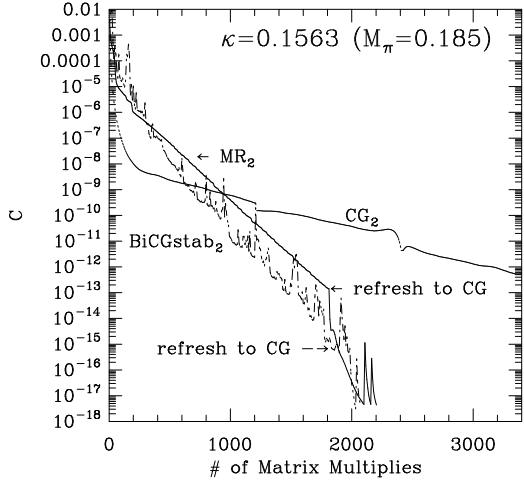}}
\figcaption{Comparison of
$CG_2$, $MR_2$ and $\bicgstab_2$ algorithms. The breaks in the
convergence of $CG_2$ are due to refreshing the remainder, however,
refreshing does not affect the aymptotic convergence rate. Data at all
values of quark mass show that $CG_2$ is not competitive with $MR_2$
or $\bicgstab_2$; however, switching to $CG$ in the final stages of
$MR$ or $\bicgstab$ algorithms enhances their convergence and makes
them comparable.}
\label{f2_compPC2k3}
\end{figure}

The $CG$ method for a non-hermitian matrix is 
\begin{eqnarray}
do\ i=1,&\omit\span\ until\_converged \nonumber \\
    \alpha  &{}={}&  {| M^\dagger r_{i-1}|^2 / | M g_{i-1}|^2}  \nonumber \\
    \chi_i  &{}={}&  \chi_{i-1} - \alpha g_{i-1} \nonumber \\
     r_i    &{}={}&  r_{i-1}   - \alpha M g_{i-1} \nonumber \\
    \beta   &{}={}&  {|M^\dagger r_i|^2 / | M^\dagger r_{i-1}|^2} \nonumber \\
\noalign{\penalty-9000}
     g_i    &{}={}&  \beta g_{i-1}   + M^\dagger r_i \nonumber \\
	    &\omit\span test\ convergence \forcefill \nonumber \\
enddo\forcefill&&
\label{eq:CG}
\end{eqnarray}
where both $\alpha, \beta$ are complex numbers and we set $g_0 =
M^\dagger r_0$. For $M$ we use the second order preconditioned
(non-hermitian) matrix defined in \ref{eq:MRpre}. The $CG$ algorithm
shows two distinct rates of convergence, a fast initial drop, and a
slower asymptotic rate that is relevant for light quarks and makes
$CG$ less competitive than $MR$ or \bicgstab. The only cases where we
have seen some gain due to the fast initial convergence behavior of
$CG$ is on switching from $MR$ or \bicgstab\ at late stages of
convergence, and even then only for light quarks. This is
\vadjust{\penalty10000} exemplified
in Fig.~\ref{f2_compPC2k3}.

\subsection{\boldmath \bicg}

\begin{figure}[t]
{\epsfxsize=0.9\hsize\epsfbox{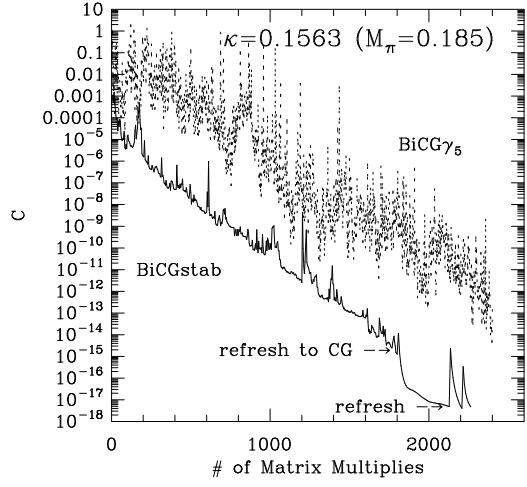}}
\figcaption{Comparison of $\bicg$ and $\bicgstab_1$ algorithms. Both solve 
for $\chi$ with respect to the un-preconditioned matrix $1+\kappa D$.}
\label{f4_BiCG5k3}
\end{figure}

The $\bicg$ algorithm exploits the property
$\gamma_5M\gamma_5=M^\dagger$ of Dirac fermions to yield a stabilized
version of $CG$ algorithm for the matrix $M$ itself \cite{rbicg5}
\begin{eqnarray}
do\ i=1,&\omit\span\ until\_converged \nonumber \\
    \alpha  &{}={}& {\langle g_{i-1}, \gamma_5 g_{i-1} \rangle / 
                \langle g_{i-1}, \gamma_5 M g_{i-1} \rangle}  \nonumber \\
    \chi_i  &{}={}&  \chi_{i-1} - \alpha g_{i-1} \nonumber \\
     r_i    &{}={}&  r_{i-1}   - \alpha M g_{i-1} \nonumber \\
    \beta   &{}={}& {\langle r_i, \gamma_5 r_i \rangle / 
                \langle r_{i-1}, \gamma_5 r_{i-1} \rangle}  \nonumber \\
\noalign{\penalty-9000}
     g_i    &{}={}&  \beta g_{i-1}   + r_i \nonumber \\
	    &\omit\span test\ convergence\forcefill \nonumber \\
enddo\forcefill&&
\label{eq:biCG}
\end{eqnarray}
where $\alpha$ and $\beta$ are real and $g_0 = r_0$. This algorithm,
in our opinion, has two drawbacks as shown in Fig.~\ref{f4_BiCG5k3};
it is sensitive to the initial $\chi_0$, for example the choice
$\chi_0 = 0$ is bad as the algorithm actually blows up.  A second
pecularity of this algorithm is that the convergence shows very large
fluctuations with a period of $\sim 10$ iterations. Overall, we do not
see any advantage to using it since the asymptotic convergence rate is
no better than $MR$ or \bicgstab.

Forcrand and Borici have raised the possibility that the problems could be 
due to round-off errors in the accumulations.  For this purpose we 
ran our tests with dot products done in 64 bit precision. This change 
did not improve the convergence. 

\subsection{\boldmath \bicgstab}

The steps in the $\bicgstab$ algorithm are \cite{rfrommer94}
\begin{eqnarray}
do\ i=1,&\omit\span\ until\_converged \nonumber \\
    \beta   &{}={}&  {\langle \hat r_0,     r_{i-1} \rangle / 
                 \langle \hat r_0, r_{i-2} \rangle} \ %
	         {\alpha \over \omega_{i-1}}  \nonumber \\
     p_i    &{}={}&  r_{i-1} + \beta (p_{i-1} - \omega v_{i-1}) \nonumber \\
     v_i    &{}={}&  M p_i \nonumber \\
    \alpha  &{}={}&  {\langle \hat r_0, r_{i-1} \rangle / {\langle \hat r_0,  v_i \rangle} }\nonumber \\
     s_i    &{}={}&  r_{i-1} - \alpha v_i \nonumber \\
     t_i    &{}={}&  M s_i \nonumber \\
    \omega_i&{}={}& {\langle t_i, s_i \rangle / {\langle t_i,t_i\rangle}} \nonumber \\
    \chi_i  &{}={}&  \chi_{i-1} + \omega s_i + \alpha p_i \nonumber \\
\noalign{\penalty-9000}
     r_i    &{}={}&  s_i  - \omega t_i \nonumber \\
	    &\omit\span test\ convergence \forcefill\nonumber \\
enddo\forcefill&&
\label{eq:biCGstab}
\end{eqnarray}
where $\alpha, \beta, \omega$ are complex numbers initialized to
$(1,0)$, $v_0 = p_0 = 0$, and $\hat r_0 = r_0 = \phi - M\chi_0$. We have tested this
algorithm for two cases of $M$, the original Dirac matrix and the
second-order preconditioned version and label these as $\bicgstab_1$
and $\bicgstab_2$ respectively.

\begin{figure}[t]
{\epsfxsize=0.9\hsize\epsfbox{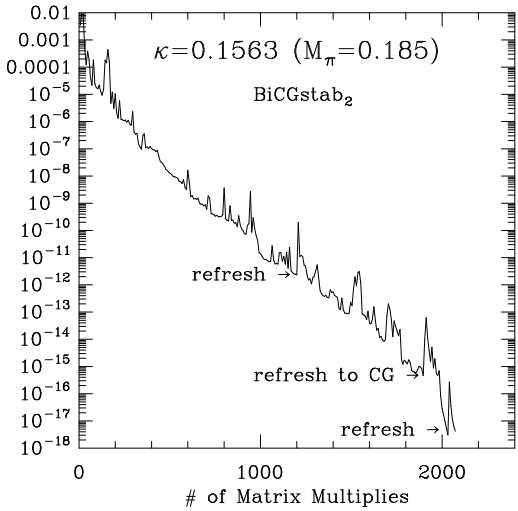}}
\vskip -0.1 truein
\figcaption{Performance details of $\bicgstab_2$ algorithm. The 
effects of refreshing the remainder and switching to $CG$ are highlighted.}
\label{f3_BiCGstabk3}
\vskip -0.1 truein
\end{figure}

The sharp spikes at count 600 and 1200 in Fig.~\ref{f3_BiCGstabk3} result from
refreshing the remainder. The data show that the convergence rate is
not affected by these spikes. The sharp drop at count 1900 is caused by 
a switch to $CG$ and is apparent only for the smallest quark mass. 

\section{Clover action}

\begin{figure}[t]
{\epsfxsize=0.9\hsize\epsfbox{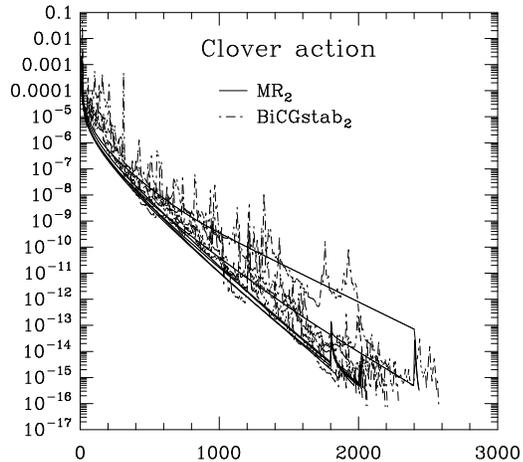}}
\vskip -0.1 truein
\figcaption{Comparison of $MR_2$ (solid lines) and $\bicgstab_2$ (dotted) 
algorithms. In both cases data are shown for six distinct lattices.}
\label{f5_clover}
\vskip -0.1 truein
\end{figure}

In Fig.~\ref{f5_clover} we show the convergence behavior of $MR_2$
and $BiCGstab_2$ algorithms when inverting the tadpole improved clover
action at $\beta=6.0$. The dependence on gauge configurations is
exhibited by using 6 distinct lattices and the lightest
quark mass ($m_\pi = 0.185$).  Our conclusion is that the performance of 
both algorithms is comparable. If one is concerned about the 
stability of $MR_2$, then one can use $MR_4$. The overhead of the clover 
term is $\approx 15\%$. 

\centerline{\bf Acknowledgements}

We thank A. Borici, C. Davies, and P. de Forcrand for discussions on
the distribution of eigenvalues and for providing data.  We thank
S.~Sharpe for comments on this work. R.G. is very grateful to Klaus
Schilling for organizing a very informative meeting and for inviting
him. We acknowledge the support of the Advanced Computing Lab at Los
Alamos for these simulations.


\end{document}

%% file: t_algops.tex
\newcommand\ce[1]{\multicolumn{#1}{|c|}}
\setlength{\tabcolsep}{2.6pt}
\begin{tabular}{|l|c|l|l|c|}
\hline
Algorithm   &Mults     &saxpy$+$ &dot$+$   & Storage   \cr
            &by $M$    &caxpy    &cdot     &           \cr
\hline
$MR_2      $ &    $2$   &  $0+2 $ & $1+1$  & 3 \cr
$MR_4      $ &    $4$   &  $0+2 $ & $1+1$  & 3 \cr
$CG_2      $ &    $4$   &  $1+2 $ & $2+0$  & 4 \cr
$\bicg     $ &    $4$   &  $3+0 $ & $0+2$  & 3 \cr
$\bicgstab_1$&    $2$   &  $0+6 $ & $1+3$  & 7 \cr
$\bicgstab_2$&    $4$   &  $0+6 $ & $1+3$  & 7 \cr
\hline
\end{tabular}

%% file: paper.bbl
\begin{thebibliography}{19}
\ifx\href\undefined\def\href#1#2{{#2}}\fi
\def\spireshome{http://www.slac.stanford.edu/cgi-bin/spiface/find/hep/www?FORMA\
T=WWW&}
{\catcode`\%=12
\xdef\spiresjournal#1#2#3{\noexpand\href{\spireshome
                          rawcmd=find+journal+#1%2C+#2%2C+#3}}
\xdef\spireseprint#1#2{\noexpand\href{\spireshome rawcmd=find+eprint+#1%2F#2}}
\xdef\spiresreport#1{\noexpand\href{\spireshome rawcmd=find+rept+#1}}
}
\def\eprint#1#2{\spireseprint{#1}{#2}{#1/#2}}
\def\report#1{\spiresreport{#1}{#1}}
\sfcode`\.=1000
\let\bf=\relax
\let\it=\relax
\def\NP#1 (#2) #3{\spiresjournal{Nucl.+Phys.}{#1}{#3}%
                   {{\it Nucl. Phys.} {\bf #1} (#2) #3}}
\def\NPB#1{\NP{B#1}}
\def\PRL#1 (#2) #3{\spiresjournal{Phys.+Rev.+Lett.}{#1}{#3}%
           {{\it Phys. Rev. Lett.} {\bf #1} (#2) #3}}
\def\PR#1 (#2) #3{\spiresjournal{Phys.+Rev.}{#1}{#3}%
           {{\it Phys. Rev.} {\bf #1} (#2) #3}}
\def\PRD#1{\PR{D#1}}
\def\PL#1 (#2) #3{\spiresjournal{Phys.+Lett.}{#1}{#3}%
          {{\it Phys. Lett.} {\bf #1} (#2) #3}}
\def\PLB#1{\PL{B#1}}
\def\IJMP#1 (#2) #3{\spiresjournal{Int.+J.+Mod.+Phys.}{#1}{#3}%
          {{\it Int. J. Mod. Phys.}{\bf #1} (#2) #3}}
%
%
\bibitem{rLANLhm95} T.~Bhattacharya, \etal, \eprint{hep-lat}{9512021}.
\bibitem{rALGlanl} R. Gupta, \etal, \PRD{40} (1989) {2072}; and 
 \PRD{44} (1991) {3272}.
\bibitem{rBorici}  A. Borici and P. deForcrand, private communications. 
\bibitem{rbicg5} P. de Forcrand, \eprint{hep-lat}{9509082}.
\bibitem{rfrommer94} Frommer, \etal, Int. \IJMP C5 (1994) {1073}.
\end{thebibliography}
